\documentclass[preprint]{aastex631}

\newcommand{\tess}{\emph{TESS}}
\newcommand{\toi}{TOI\,451}
\newcommand{\pyaneti}{\href{https://github.com/oscaribv/pyaneti}{\texttt{pyaneti}}}

\newcommand{\sradius}[1][$R_{\odot}$]{ $0.879 \pm 0.032 $ #1}
\newcommand{\Tzerob}[1][days] {$1410.9896 _{ - 0.0029 } ^ { + 0.0032 }$~#1} 
\newcommand{\Pb}[1][days]   {$1.8587028 _{ - 10e-06 } ^ { + 08e-06 }$~#1} 
 
\newcommand{\rrb}[1][ ]   {$0.01974 \pm 0.00068 $~#1} 
\newcommand{\rpb}[1][$R_{\oplus}$]   {$1.892 _{ - 0.094 } ^ { + 0.096 }$~#1}

\newcommand{\Tzeroc}[1][days]   {$1411.7982 _{ - 0.0020 } ^ { + 0.0022 }$~#1} 
\newcommand{\Pc}[1][days]   {$9.192453 _{ - 3.3e-05 } ^ { + 4.1e-05 }$~#1} 
 
\newcommand{\rrc}[1][ ]   {$0.03217 _{ - 0.00086 } ^ { + 0.00083 }$~#1} 
\newcommand{\rpc}[1][$R_{\oplus}$]   {$3.08 \pm 0.14$~#1}

\newcommand{\Tzerod}[1][days]   {$1416.63407 _{ - 0.00100 } ^ { + 0.00096 }$~#1}
\newcommand{\Pd}[1][days]   {$16.364932 _{ - 3.5e-05 } ^ { + 3.6e-05 }$~#1} 
 
\newcommand{\rrd}[1][ ]   {$0.04296 _{ - 0.00078 } ^ { + 0.00080 }$~#1} 
\newcommand{\rpd}[1][$R_{\oplus}$]   {$4.12 \pm 0.17 $~#1}

\shorttitle{TOI-451 update}
\shortauthors{Barrag\'an et al.}

\begin{document}

\title{TESS re-observes the young multi-planet system TOI-451: \\
refined ephemeris and activity evolution}

\correspondingauthor{Oscar Barrag\'an}
\email{oscar.barragan@physics.ox.ac.uk}

\author[0000-0003-0563-0493]{Oscar Barrag\'an}
\affiliation{Department of Physics, University of Oxford, OX13RH, Oxford, UK}

\author[0000-0003-1453-0574]{Suzanne Aigrain}
\affiliation{Department of Physics, University of Oxford, OX13RH, Oxford, UK}

\author[0000-0003-2851-3070]{Edward Gillen}
\affiliation{Astronomy Unit, Queen Mary University of London, Mile End Road, London E14NS, UK}
\affiliation{Astrophysics Group, Cavendish Laboratory, J.J. Thomson Avenue, Cambridge CB30HE, UK}
\affiliation{Winton Fellow}

\author{Fernando Guti\'errez-Canales}
\affiliation{Departamento de Astronom\'ia, Universidad de Guanajuato, Callej\'on de Jalisco, 36023, M\'exico}



\begin{abstract}
We present a new analysis of the light curve of the young planet-hosting star \toi\ in the light of new observations from \tess\ Cycle 3. Our joint analysis of the transits of all three planets, using all available \tess\ data, results in an improved ephemeris for \toi\,b and \toi\,c, which will help to plan follow-up observations. The updated mid-transit times are $\textrm{BJD}-2,457\,000=$\Tzerob[], \Tzeroc[], and \Tzerod[] for
 \toi\ b, c, and d, respectively, and the periods are \Pb[], \Pc[], and \Pd[] days. We also model the out-of-transit light curve using a Gaussian Process with a quasi-periodic kernel, and infer a change in the properties of the active regions on the surface of \toi\ between \tess\ Cycles 1 and 3.
\end{abstract}



\section*{} 

Using data collected by NASA's Transiting Exoplanet Survey Satellite \citep[\tess;][]{Ricker2015} during its first year of operations, combined with ground-based follow-up observations, \citet[][hereafter N21]{Newton2021} discovered a system of three transiting planets orbiting the star \toi, a member of the young (120 Myr) Pisces–Eridanus stream. As the host star is relatively bright ($V=11.02$), it is amenable to further follow-up, including Radial Velocity (RV) observations to measure the planetary masses, and transmission spectroscopy to characterise the planets' atmospheres. Further characterisation of a system like \toi\ is expected to provide valuable constraints for models of planet formation and early evolution, but depends critically on our ability to predict the times of future transits accurately, and model the signals arising from active regions on the star's surface, which are important for both RV data and transmission spectra.

\tess\ first observed \toi\ (TIC 257605131) in Cycle 1 during sectors 4 and 5 (from 2018-Oct-18 to 2018-Dec-11). These observations, together with additional ground- and space-based photometry, were used by N21 to discover and validate the three transiting planets which are known around \toi\ to date. Two years later, \tess\ re-visited \toi\ in its extended mission during its Cycle 3 in sector 31 (from 2020-Oct-21 to 2020-Nov-19). In this Note we analyse the \tess\ data from Cycles 1 and 3 jointly. 

We downloaded the \tess\ light curves for the three sectors from the \href{https://archive.stsci.edu/missions-and-data/tess}{Mikulski Archive for Space Telescopes}. 
We use the Pre-search Data Conditioning Simple Aperture Photometry (PDC SAP) light curve as the starting point for our analysis. This light curve is corrected for known instrumental effects at the pixel level, as well as for common-mode systematics. 

\begin{figure}
    \centering
    \includegraphics[width=1\textwidth]{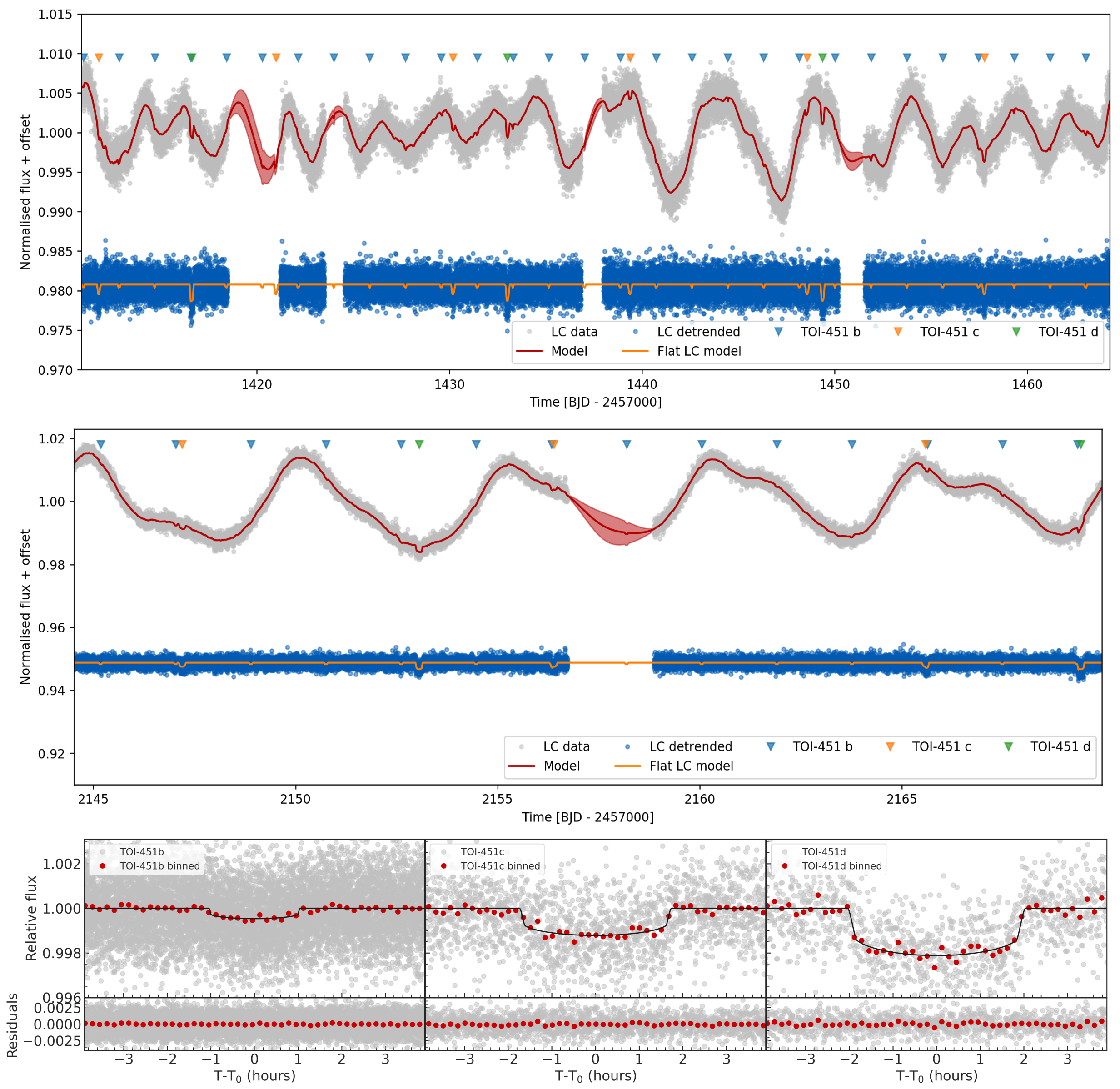}
    \caption{Top panel: PDC-SAP light curve (grey) along with best-fit GP$+$transit model (red), for \tess\ sectors 4 \& 5. Also shown, with a vertical offset for clarity, are the detrended light curve (blue) and the transit-only model (orange). Middel panel: same as top panel, but for sector 31. Bottom row: detrended light curves phase-folded at the period of each planet (individual data points in grey, phase-binned data in red) together with the best-fit transit model (black line), with the residuals in the lower inset.}
    \label{fig}
\end{figure}

Before modelling the transits, we detrend the light curve with \href{https://github.com/oscaribv/citlalicue}{\texttt{citlalicue}}, which uses a quasi-periodic Gaussian Process (GP) implemented in \texttt{george} \citep[][]{george} to model the activity-induced variability in the out-of-transit data, together with \texttt{pytransit} \citep[][]{pytransit} to predict the times at which transits occur. We mask out all the transits when fitting the GP, and remove other outliers by clipping at $\pm 5$-sigma. We then divide by the GP model to obtain a detrended light curve containing transits only. Because the Cycle 1 and 3 observations were taken two years apart, we detrended them separately. Figure \ref{fig} shows the PDC SAP light curve, together with the GP model, and the detrended light curve, phase-folded at the period of each planet and zoomed in on the transits. We return to the GP model at the end of this note when discussing activity.

We then extracted sections of the detrended light curve around each transit, and modelled them using  \texttt{\pyaneti} \citep{pyaneti}. We model the transits of all a three planets simultaneously, assuming circular orbits. For each planet, we vary the time of mid-transit $T_0$, orbital period $P$, impact parameter $b$, and scaled planet radius, $R_{\rm p}/R_\star$. We also vary the stellar density, and convert it to scaled semi-major axis for each planet using Kepler's third law. Finally, we model for the stellar limb darkening following a quadratic limb darkening model. 
We used wide uniform priors for all the parameters. This results in the following estimates for the planet parameters (b, c, and d respectively on each line):
\begin{itemize}
    \item $T_0 (\textrm{BJD}-2\,457\,000) = $\Tzerob[], \Tzeroc[], \Tzerod[];
    \item $P (\textrm{days}) = $\Pb[], \Pc[], \Pd[];
    \item $R_{\rm p}/R_\star = $\rrb, \rrc, \rrd;
    \item $R_{\rm p} (R_\oplus) =$ \rpb[], \rpc[], \rpd[].
\end{itemize}
where we converted from $R_{\rm p}/R_\star$ to $R_{\rm p}$ using the stellar radius of \sradius\ given by N21.




All our estimates are consistent with the values reported by N21, but our ephemeris for planets b and c are significantly improved, by a factor $\sim 3$ and $\sim 2$, respectively.
We do not improve on the ephemeris for \toi\,d as Sector 31 includes only two new transits of this planet, and N21 combined data from \tess\ Sector 4 and 5 with five additional transits observed with other facilities, which we do not analyse here. The best-fit transit model for each planet is shown in the bottom row of Figure~\ref{fig}.


As expected for such a young star, \toi\ is active, and displays significant out-of-transit variability due to the rotational modulation and evolution of the active regions on the stellar surface, which is clearly  visible in the top panel of Figure~\ref{fig}. It is interesting to note that this variability looks qualitatively different in \tess\ Cycle 1 and 3. Compared to sectors 4 \& 5, the stellar variability in sector 31 has a larger amplitude and appears to be more coherent. Cycle 1 also showed hints of a `beat pattern', which is not seen in sector 31. Such beat patterns can arise when two or more active regions are present on the stellar surface, and evolve or rotate at different rates. 

To quantify these differences, we explore the range of parameters of our quasi-periodic GP model which are compatible with the data in each Cycle. To speed up this process, since we are only interested variations on timescales of a day or more, we binned the data in 6 hour bins. We used \texttt{\pyaneti} to model the light curves using a GP with the Quasi-Periodic kernel:
\begin{equation}
    \gamma_{\rm QP}(t_i,t_j) = A^2 \exp 
    \left[
    - \frac{\sin^2\left[\pi \left(t_i - t_j \right)/P_{\rm GP}\right]}{2 \lambda_{\rm p}^2}
    - \frac{\left(t_i - t_j\right)^2}{2\lambda_{\rm e}^2}
    \right]
    \label{eq:qp}
    ,
\end{equation}
where $P_{\rm GP}$ is the period of the activity signal, $\lambda_{\rm p}$ is the length scale of the periodic component, and $\lambda_{\rm e}$ is the long term evolution timescale. While $P_{\rm GP}$ and $\lambda_{\rm e}$ have units of days, $\lambda_{\rm p}$ is dimensionless.

The results of this analysis are as follows (each line reports the value for sectors 4 \& 5 first, then the value for sector 31):
\begin{itemize}
    \item $A = 2.8 _{ - 0.3 } ^ { + 0.4 } \times 10^{-3}$, $9.8_{-1.9}^{+2.5} \times 10^{-3}$, 
    \item $\lambda_{\rm e} (\textrm{days})= 4.93 _{ - 0.32 } ^ { + 0.33 }$, $8.17 _{ - 0.72 } ^ { + 0.80 }$,
    \item $ \lambda_{\rm p} = 0.359 _{ - 0.022 } ^ { + 0.025 }$, $ 0.699 _{ - 0.093 } ^ { + 0.136 }$
    \item $ P_{\rm GP}  (\textrm{days}) = 5.127 _{ - 0.056 } ^ { + 0.064 }$, $5.184 _{ - 0.058 } ^ { + 0.052 }$.
\end{itemize}

As expected, the GP period is consistent between both analyses, and they agree with the stellar rotation period derived by N21. However, the remaining hyper-parameters are not consistent between the two data sets. Specifically, the GP amplitude $A$, the evolutionary time-scale $\lambda_{\rm e}$, and the periodic length scale $\lambda_{\rm p}$ are all significantly larger in \tess\ Cycle 3 than in Cycle 1. This confirms our qualitative visual assessment of the light curve, and indicates that the size and distribution of the active regions has evolved significantly during the two years separating the two sets of observations.

This may prove important when planning future follow-up observations of the target. In particular, we recommend attempting to obtain photometric monitoring contemporaneous with any future spectroscopic observations, and/or 
explicitly using activity indicators extracted from the spectra themselves to disentangle between stellar and planetary signals \citep[e.g.,][]{Rajpaul2015,Barragan2019}. Relying on the light curves from previous seasons to constrain the lifetime and distribution of the active regions on the stellar surface may not be appropriate for this target.

\section*{Acknowledgements}

We acknowledge the use of public TESS Alert data from pipelines at the TESS Science Office and at the TESS Science Processing Operations Center.
Resources supporting this work were provided by the NASA High-End Computing (HEC) Program through the NASA Advanced Supercomputing (NAS) Division at Ames Research Center for the production of the SPOC data products.
This work received funding from the European Research Council under the European Union’s Horizon 2020 research and innovation program (Grant agreement No. 865624).
EG acknowledges support from the David and Claudia Harding Foundation in the form of a Winton Exoplanet Fellowship.
FGC thanks the Mexican national council for science and technology (CONACYT, CVU-1005374).


\bibliography{bibs}
\bibliographystyle{aasjournal}



\end{document}